\documentstyle[aas2pp4]{article}

\begin{document}

\twocolumn

\noindent {\small ADP-AT-98-9} 

\title{Acceleration and Interaction of Ultra High Energy Cosmic Rays}

\author{ R.J. Protheroe }
\author{\it Department of Physics and Mathematical Physics\\ 
The University of Adelaide, Adelaide, Australia 5005\\}

\begin{abstract}
In this chapter I give an overview of shock acceleration,
including a discussion of the maximum energies possible and the
shape of the spectrum near cut-off, interactions of high
energy cosmic rays with, and propagation through, the background
radiation, and the resulting electron-photon cascade.  Possible
sources of the highest energy cosmic rays are discussed including
active galaxies, gamma ray bursts and topological defects.  I
argue that while the origin of the highest energy cosmic rays is
still uncertain, it is not necessary to invoke exotic models such
as emission by topological defects to explain the existing data.
It seems likely that shock acceleration at Fanaroff-Riley Class
II radio galaxies can account for the existing data.  However,
new cosmic ray data, as well as better estimates of the
extragalactic radiation fields and magnetic fields will be
necessary before we will be certain of the origin of the highest
energy particles occurring in nature.
\end{abstract}

\vspace{.2in}

\section{Introduction}

Cosmic rays with energies up to 100 TeV are thought to arise
predominantly through shock acceleration by supernova remnants
(SNR) in our Galaxy (Lagage \& Cesarsky 1983)\markcite{Lag83}.  A
fraction of the cosmic rays accelerated should interact within
the supernova remnant and produce $\gamma$--rays (Drury et al. 1994,
Gaisser et al. 1998, Baring et al. 1999)
\markcite{DruryAharonianVolk,GaisserProtheroeStanev98,Baring99},
and recent observations above 100 MeV by the EGRET instrument on
the Compton Gamma Ray Observatory have found $\gamma$-ray signals
associated with at least two supernova remnants -- IC~443 and
$\gamma$~Cygni (Esposito et al. 1996)\markcite{Esposito96}.
However, Brazier et al. (1996)\markcite{Brazier96} have suggested
that the $\gamma$-ray emission from IC~443 may be associated with a
pulsar within the remnant rather than the remnant itself.
Further evidence for acceleration in SNR comes from the Rossi
X-ray Timing Explorer observations of Cassiopeia A showing a
non-thermal component in the spectrum (Allen et
al. 1997)\markcite{Allen97}, and the ASCA observation of
non-thermal X--ray emission from SN~1006 (Koyama et
al. 1995)\markcite{Koyama95}.
Reynolds (1996)\markcite{Reynolds96} and Mastichiadis
(1996)\markcite{Mastichiadis96} interpret the latter as
synchrotron emission by electrons accelerated in the remnant up
to energies as high as 100 TeV, although Donea and Biermann
(1998)\markcite{DoneaBiermann98} suggest it may be bremsstrahlung
from much lower energy electrons.  The CANGAROO telescope appears
to have detected TeV $\gamma$-rays from SN1006 (Tanimori et
al. 1998)\markcite{Tanimori98}, while there has been a
disappointing lack of detections of TeV $\gamma$-rays from IC443 and
$\gamma$~Cygni.  This may be related to the matter density in
which the SNR shocks propagate (Baring et al. 1999), higher
densities yielding lower cut-off energies for IC443 and
$\gamma$~Cygni.

Acceleration to somewhat higher energies than 100 TeV may be
possible (Markiewicz et al. 1990)\markcite{Mar90}, but probably
not to high enough energies to explain the smooth extension of
the spectrum to 1 EeV.  Several explanations for the origin of
the cosmic rays in this energy range have been suggested:
reacceleration of the supernova component while still inside the
remnant (Axford 1991)\markcite{Axf91}; by several supernovae
exploding into a region evacuated by a pre-supernova star (Ip and
Axford 1991)\markcite{Ip91}; or acceleration in shocks inside the
strong winds from hot stars or groups of hot stars (Biermann and
Cassellini 1993)\markcite{Bie93}.  Very recently, an analysis of
arrival direction data from AGASA (Hayashida et al.
1998\markcite{Hayashida98}) shows excesses in the directions of
the Galactic Centre and the Cygnus region which could not easily
be explained by charged particle propagation from these sources.
If indeed the excess is due to these sources it may provide
some evidence for a component of the cosmic rays at 1 EeV being
neutrons.  At 5 EeV the spectral slope changes, and there is
evidence for a lightening in composition (Bird et al. 1994,
Dawson et al. 1998)\markcite{Bir94,Dawson98} and it is likely
this marks a change from galactic cosmic rays to extragalactic
cosmic rays being dominant.

An alternative, although less popular, scenario is that almost
all cosmic ray nuclei are of extragalactic origin (e.g. Brecher
and Burbidge 1972\markcite{BrecherBurbidge72}, and references
therein).  In any case, whether or not the lower energy cosmic
rays are indeed galactic, at the highest energies (above $\sim
10^{19}$~eV) it is very probably extragalactic.

The cosmic ray air shower events with the highest energies so far
detected have energies of $2 \times 10^{11}$ GeV (Hayashida et
al. 1994)\markcite{Hay94} and $3 \times 10^{11}$ GeV (Bird et
al. 1995)\markcite{Bir95}, and recent results from AGASA have
shown there to be a continuous spectrum between $10^{11}$ GeV and
$3 \times 10^{11}$ GeV (Takeda et al. 1998)\markcite{Takeda98}).
The question of the origin of these cosmic rays having energy
significantly above $10^{11}$ GeV is complicated by propagation
of such energetic particles through the Universe.  Nucleons
interact with the cosmic background radiation fields, losing
energy by Bethe-Heitler pair production, or interacting by pion
photoproduction, and in the latter case may emerge as either
protons or neutrons with reduced energy.  The threshold for pion
photoproduction on the microwave background is $\sim 2 \times
10^{10}$ GeV, and at $3 \times 10^{11}$ GeV the energy-loss
distance is about 20 Mpc.  Propagation of cosmic rays over
substantially larger distances gives rise to a cut-off in the
spectrum at $\sim 10^{11}$ GeV as was first shown by Greisen
(1966)\markcite{Gre66}, and Zatsepin and Kuz'min
(1966)\markcite{Zat66}, the ``GZK cut-off'', and a corresponding
pile-up at slightly lower energy (Hill and Schramm 1985,
Berezinsky and Grigor'eva 1988)\markcite{Hil85,Ber88}.  These
processes occur not only during propagation, but also during
acceleration, and may actually limit the maximum energies
particles can achieve.

In this chapter I give an overview of shock acceleration,
describe interactions of high energy protons and nuclei with
radiation, discuss maximum energies obtainable during
acceleration, and the shape of the spectrum near maximum energy,
outline propagation of cosmic rays through the background
radiation and the consequent electron-photon cascading, and
finally discuss conventional and exotic models of the highest
energy cosmic rays.

%%%%%%%%%%%%%%%%%%%%%%%%%%%%%%%%%%%%%%%%%%%%%%%%%%%%%%%%%%%%%%%%%%%%%%%%%%%%%%%

\section{Interactions of High Energy Cosmic Rays}

Interactions of cosmic rays with radiation 
are important both during acceleration when the resulting energy
losses compete with energy gains by, for example, shock
acceleration, and during propagation from the acceleration region
to the observer.  For ultra-high energy (UHE) cosmic rays, the
most important processes are pion photoproduction and
Bethe-Heitler pair production both on the microwave background,
and synchrotron radiation.  In the case of nuclei,
photodisintegration on the microwave background is important.  In
this section, I shall describe how to calculate the mean free
path for such interactions, and briefly discuss how to simulate
the interactions using the Monte Carlo method.

\subsection{Nucleons}

The mean interaction length, $x_{p \gamma}$, of a proton of
energy $E$ is given by,
\begin{equation}
        {1 \over x_{p \gamma}(E)}= {1 \over 8 \beta E^2}
\int_{\varepsilon_{\rm min}(E)}^
        {\infty} \frac{n(\varepsilon)}{\varepsilon^2} 
	\int_{s_{\rm min}}^{s_{\rm max}(\varepsilon,E)} \hspace{-3mm}
        \sigma(s)(s-m_p^2 c^4)ds d\varepsilon,
        \label{eq:mpl}
\end{equation}
where $n(\varepsilon)$ is the differential photon number density
of photons of energy $\varepsilon$, and $\sigma(s)$ is the
appropriate total cross section for the process in question for a
centre of momentum (CM) frame energy squared, $s$, given by
\begin{equation}
s=m_p^2 c^4 + 2 \varepsilon E(1 - \beta \cos \theta)
\label{eq:s}
\end{equation}
where $\theta$ is the angle between the directions of the 
proton and photon,
and $\beta c$ is the proton's velocity.

For pion photoproduction  
\begin{equation}
s_{\rm min} = (m_pc^2+m_{\pi}c^2)^2 \approx 1.16 \; \rm GeV^2,
\end{equation}
and
\begin{equation}
\varepsilon_{\rm min}= {m_{\pi}c^2(m_{\pi}c^2+2m_pc^2) \over 2E(1+\beta)}
\approx {m_{\pi}c^2(m_{\pi}c^2+2m_pc^2) \over 4E}.
\end{equation}
For photon-proton pair-production the threshold is somewhat lower,
\begin{equation}
s_{\rm min}= (m_pc^2 + 2 m_ec^2)^2 \approx0.882 \; \rm GeV^2,
\end{equation}
and
\begin{equation}
\varepsilon_{\rm min}\approx m_ec^2(m_ec^2+m_pc^2)/E.
\end{equation}
For both processes,
\begin{equation}
s_{\rm max}(\varepsilon,E) = m_p^2c^4+2\varepsilon E(1+\beta)
\approx m_p^2c^4+4\varepsilon E,
\end{equation} 
and $s_{\rm max}(\varepsilon,E)$ corresponds to a head-on collision
of a proton of energy $E$ and a photon of energy $\varepsilon$.

Examination of the integrand in Equation \ref{eq:mpl} shows that the 
energy of the soft photon interacting with a proton 
of energy $E$ is distributed as
\begin{equation}
p(\varepsilon) = \frac{x_{p \gamma}(E)n(\varepsilon)}{8 \beta E^2 \varepsilon^2}
        \Phi(s_{\rm max}(\varepsilon,E))
\end{equation}
in the range $\varepsilon_{\rm min} \leq \varepsilon \leq \infty$ where 
\begin{equation}
        \Phi(s_{\rm max})=\int_{s_{\rm min}}^{s_{\rm max}}\sigma(s)
        (s-m_p^2c^4)ds.
\end{equation}
Similarly, examination of the integrand in Equation \ref{eq:mpl} shows that the
square of the
total CM frame energy is distributed as
\begin{equation}
        p(s) =\frac{\sigma(s)(s-m_p^2c^4)}{\Phi(s_{\rm max})},
\end{equation}
in the range $s_{\rm min}\leq s\leq s_{\rm max}$.

The Monte Carlo rejection technique can be used to sample
$\varepsilon$ and $s$ respectively from the two distributions,
and Equation \ref{eq:s} is used to find $\theta$.  One then
Lorentz transforms the interacting particles to the frame in
which the interaction is treated (usually the proton rest frame),
and samples momenta of particles produced in the interaction from
the appropriate differential cross section by the rejection
method.  The energies of produced particles are then Lorentz
transformed to the laboratory frame, and the final energy of the
proton is obtained by requiring energy conservation.  In this
procedure, it is not always possible to achieve exact
conservation of both momentum and energy while sampling particles
from inclusive differential cross sections (e.g. multiple pion
production well above threshold), and the momentum of the last
particle sampled is therefore adjusted to minimize the error.

\begin{figure}[htb]
\plotone{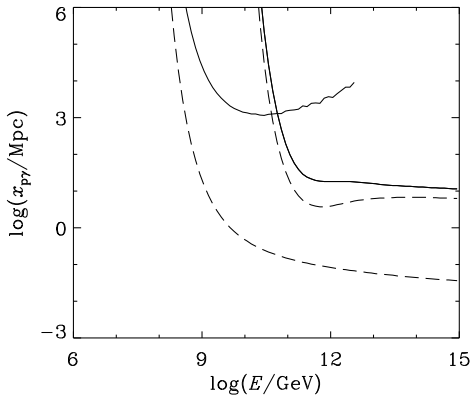}
\caption{Mean interaction length (dashed lines) and energy-loss 
distance (solid lines), $E/(dE/dx)$, for proton-photon pair-production
and pion-production in the microwave background
(lower and higher energy curves respectively). 
(From Protheroe and Johnson 1995)\protect \markcite{ProtheroeJohnson95}.
\label{fig:pgpiee3k_xloss}}
\end{figure}

The mean interaction lengths for both processes,
$x_{p\gamma}(E)$, are obtained from Equation \ref{eq:mpl} for
interactions in the microwave background and are plotted as
dashed lines in Fig. \ref{fig:pgpiee3k_xloss}.  Dividing by the
inelasticity, $\kappa(E)$, one obtains the energy-loss distances
for the two processes,
\begin{equation}
{E \over dE/dx} = {x_{p \gamma}(E) \over \kappa(E)}.
\end{equation}

%%%%%%%%%%%%%%%%%%%%%%%%%%%%%%%%%%%%%%%%%%%%%%%%%%%%%%%%%%%%%%%%%%%%%%%%%%%%%%%
 
\subsection{Nuclei}

In the case of nuclei the situation is a little more complicated.
The threshold condition for Bethe-Heitler pair production can be
expressed as
\begin{equation}
\gamma > {m_e c^2 \over \varepsilon} \left( 1 + {m_e \over Am_p} \right),
\end{equation}
and the threshold condition for pion photoproduction can be
expressed as
\begin{equation}
\gamma > {m_\pi c^2 \over 2 \varepsilon} \left( {1 + {m_\pi \over
2A m_p}} \right).
\end{equation}
Since $\gamma = E / A m_p c^2$, where $A$ is the mass number, we
will need to shift both energy-loss distance curves in
Fig.~\ref{fig:pgpiee3k_xloss} to higher energies by a factor of
$A$.  We shall also need to shift the curves up or down as
discussed below.

For Bethe-Heitler pair production the energy lost by a nucleus in
each collision near threshold is approximately $\Delta E \approx
\gamma 2 m_e c^2$.  Hence the inelasticity is
\begin{equation}
K \equiv { \Delta E \over E} \approx {2m_e \over A m_p},
\end{equation}
and is a factor of $A$ lower than for protons.  On the other
hand, the cross section goes like $Z^2$, so the overall shift is
down (to lower energy-loss distance) by $Z^2/A$.  For example,
for iron nuclei the energy loss distance for pair production is
reduced by a factor $26^2/56 \approx 12.1$.

\begin{figure}[htb]
\plotone{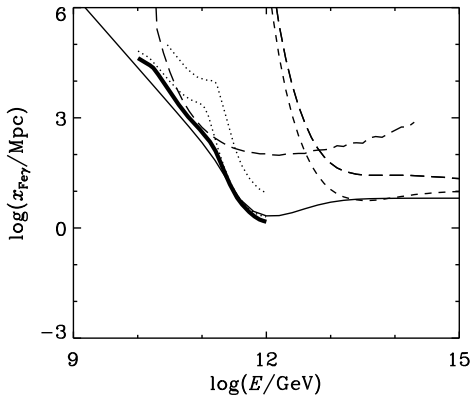}
\caption{Energy-loss distance of Fe-nuclei in the CMBR for
pair-production (leftmost long dashed line) and pion
photoproduction (rightmost long dashed line), and mean
interaction length for pion photoproduction multiplied by 56
(short dashed line) are obtained from curves in \protect
Fig.~1.  The photodisintegration distances
given by Stecker and Salamon~(1999) for loss of one nucleon (lower dotted
curve) and two nucleons (upper dotted line) are shown together
with the total loss distance estimated by Stecker and Salamon
(1999).  The thin full curve shows an estimate over a larger
range of energy (Protheroe, unpublished) of the total loss
distance based on photodisintegration cross sections of Karakula
and Tkaczyk (1993).
\label{fig:fepiee_xloss}}
\end{figure}

For pion production the energy lost by a nucleus in each
collision near threshold is approximately $\Delta E \approx
\gamma m_\pi c^2$, and so, as for pair production, the
inelasticity is factor $A$ lower than for protons.  The cross
section increases approximately as $A^{0.9}$ giving an overall
increase in the energy loss distance for pion production of a
factor $\sim A^{0.1} \approx 1.5$ for iron nuclei.
The energy loss distances for pair production and pion
photoproduction are shown for iron nuclei in
Fig.~\ref{fig:fepiee_xloss}.  

Photodisintegration is very important and has been considered in
detail by Tkaczyk et al. (1975)\markcite{Tka75}, Puget et
al. (1976)\markcite{Pug76}, Karakula and Tkaczyk (1993)
\markcite{KarakulaTkaczyk93}, Epele and Roulet (1998)
\markcite{EpeleRoulet98} and Stecker and Salamon (1999).
\markcite{SteckerSalamon99}  The photodisintegration distance
defined by $A/(dA/dx)$ taken from Stecker and Salamon (1999)
\markcite{SteckerSalamon99} is shown in
Fig.~\ref{fig:fepiee_xloss} together with an estimate made over a
larger range of energy by Protheroe (unpublished) of the total
loss distance based on photodisintegration cross sections of
Karakula and Tkaczyk (1993)\markcite{KarakulaTkaczyk93}.  Since
iron nuclei will be fragmented during pion photoproduction, the
photodisintegration distance of Fe at high energies should be
consistent with $\sim 56 \times$ {\em the mean free path} for
pion photoproduction by Fe at higher energies, and this is found
to be the case (see Fig.~\ref{fig:fepiee_xloss}).

%%%%%%%%%%%%%%%%%%%%%%%%%%%%%%%%%%%%%%%%%%%%%%%%%%%%%%%%%%%%%%%%%%%%%%%%%%%%%%%
%%%%%%%%%%%%%%%%%%%%%%%%%%%%%%%%%%%%%%%%%%%%%%%%%%%%%%%%%%%%%%%%%%%%%%%%%%%%%%%

\section{Cosmic Ray Acceleration}
%%%%%%%%%%%%%%%%%%%%%%%%%%%%%%%%%%%%%%%%%%%%%%%%%%%%%%%%%%%%%%%%%%%%%%%%%%%%%%%
%%%%%%%%%%%%%%%%%%%%%%%%%%%%%%%%%%%%%%%%%%%%%%%%%%%%%%%%%%%%%%%%%%%%%%%%%%%%%%%

For stochastic particle acceleration by electric fields induced by
motion of magnetic fields $B$, the rate of energy gain by relativistic
particles of charge $Ze$ can be written (in SI units)
\begin{equation}
\left. {dE \over dt} \right|_{\rm acc} = \xi Ze c^2 B
\label{eq:max_acc}
\end{equation}
where $\xi < 1$ and depends on the acceleration mechanism.  I
shall give a simple heuristic treatment of Fermi acceleration
based on that given in Gaisser's excellent book (Gaisser
1990)\markcite{Gaisser}.  I shall start with 2nd order Fermi
acceleration (Fermi's original theory) and describe how this can
be modified in the context of astrophysical shocks into the more
efficient 1st order Fermi mechanism known as shock acceleration.
More detailed and rigorous treatments are given in several review
articles (Drury 1983a, Blandford and Eichler 1987, Berezhko and Krymsky
1988)\markcite{Drury83a,BlandfordEichler87,BerezhkoKrymsky88}.  See
the review by Jones and Ellison (1991)\markcite{JonesEllison91}
on the plasma physics of shock acceleration which also includes a
brief historical review and refers to early work.

\subsection{Fermi's Original Theory}

Gas clouds in the interstellar medium have random velocities of
$\sim 15$ km/s superimposed on their regular motion around the
galaxy.  Cosmic rays gain energy on average when scattering off
these magnetized clouds.  A cosmic ray enters a cloud and
scatters off irregularities in the magnetic field which is tied
to the cloud because it is partly ionized.

\begin{figure}[htb]
\plotone{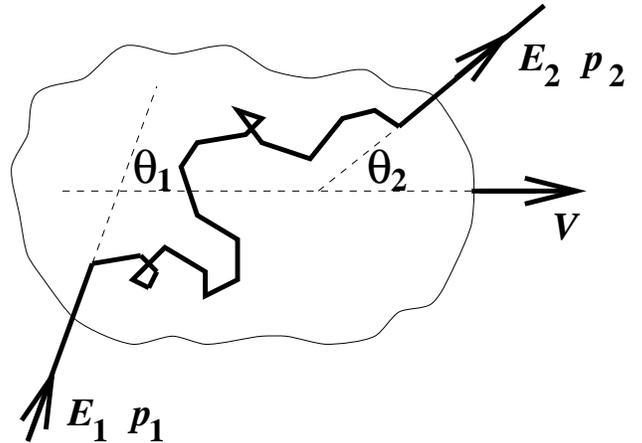}
\caption{Interaction of cosmic ray of energy $E_1$ with ``cloud''
moving with speed $V$
\label{fig:fermi_acc_orig}}
\end{figure}

In the frame of the cloud: (a) there is no change in energy
because the scattering is collisionless, and so there is elastic
scattering between the cosmic ray and the cloud as a whole which
is much more massive than the cosmic ray; (b) the cosmic ray's
direction is randomized by the scattering and it emerges from the
cloud in a random direction.

Consider a cosmic ray entering a cloud with energy $E_{1}$ and
momentum $p_{1}$ travelling in a direction making angle
$\theta_{1}$ with the cloud's direction.  After scattering inside
the cloud, it emerges with energy $E_{2}$ and momentum $p_{2}$ at
angle $\theta_{2}$ to the cloud's direction
(Fig.~\ref{fig:fermi_acc_orig}).  The energy change is obtained
by applying the Lorentz transformations between the laboratory
frame (unprimed) and the cloud frame (primed).  Transforming to
the cloud frame:
\begin{equation}
E_{1}^{\prime} = \gamma E_{1} (1 - \beta \cos \theta_{1})
\end{equation}
where $\beta = V/c$ and $\gamma = 1/\sqrt{1-\beta^{2}}$.

\noindent Transforming to the laboratory frame:
\begin{equation}
E_{2} = \gamma E_{2}^{\prime} (1 + \beta \cos \theta_{2}^{\prime}).
\end{equation}

The scattering is collisionless, being with the magnetic field.
Since the magnetic field is tied to the cloud, and the cloud is
very massive, in the cloud's rest frame there is no change in
energy, $E_{2}^{\prime} = E_{1}^{\prime}$, and hence we obtain
the fractional change in LAB-frame energy $(E_{2}-E_{1})/E_{1}$,
\begin{equation}
{\Delta E \over E} = 
{1 - \beta \cos \theta_{1} + \beta \cos \theta_{2}^{\prime}
- \beta^{2} \cos \theta_{1} \cos \theta_{2}^{\prime} \over
1 - \beta^{2}} -1.
\end{equation}

We need to obtain average values of $ \cos \theta_{1}$ and $ \cos
\theta_{2}^{\prime}$.  Inside the cloud, the cosmic ray scatters
off magnetic irregularities many times so that its direction is
randomized,
\begin{equation}
\langle \cos \theta_{2}^{\prime} \rangle =0.
\end{equation}
The average value of cos$\theta_{1}$ depends on the rate at which
cosmic rays collide with clouds at different angles.  The rate of
collision is proportional to the relative velocity between the
cloud and the particle so that the probability per unit solid
angle of having a collision at angle $\theta_{1}$ is proportional
to $(v - V \cos \theta_{1})$.  Hence, for ultrarelativistic
particles ($v=c$)
\begin{equation}
{dP \over d \Omega_{1}} \propto (1 - \beta \cos \theta_{1}),
\end{equation}
and we obtain
\begin{equation}
\langle \cos \theta_{1} \rangle = 
\int \cos \theta_{1} {dP \over d \Omega_{1}} d \Omega_{1} /
\int {dP \over d \Omega_{1}} d \Omega_{1} = - {\beta \over 3},
\end{equation} 
giving
\begin{equation}
{\langle \Delta E \rangle \over E} = 
{1 + \beta^{2}/3 \over
1 - \beta^{2}} -1 \simeq {4 \over 3} \beta^{2}
\end{equation}
since $\beta \ll 1$.

We see that $\langle \Delta E \rangle / E \propto \beta^{2}$ is
positive (energy gain), but is 2nd order in $\beta$ and because
$\beta \ll 1$ the average energy gain is very small.  This is
because there are almost as many overtaking collisions (energy
loss) as there are head-on collisions (energy gain).

%%%%%%%%%%%%%%%%%%%%%%%%%%%%%%%%%%%%%%%%%%%%%%%%%%%%%%%%%%%%%%%%%%%%%%%%%%%%

\subsection{1st Order Fermi Acceleration at SN or Other Shocks}

Fermi's original theory was modified in the 1970's (Axford, Lear
and Skadron 1977, Krymsky 1977, Bell 1978, Blandford and Ostriker
1978)\markcite{Axford77,Krymsky77,Bell78,BlandfordOstriker78} to
describe more efficient acceleration (1st order in $\beta$)
taking place at supernova shocks but is generally applicable to
strong shocks in other astrophysical contexts.  Our discussion of
shock acceleration will be of necessity brief, and omit a number
of subtleties.

Here, for simplicity, we adopt the test particle approach
(neglecting effects of cosmic ray pressure on the shock profile),
adopt a plane geometry and consider only non-relativistic
shocks.  Nevertheless, the basic concepts will be described in
sufficient detail that we can consider acceleration and
interactions of the highest energy cosmic rays, and to what
energies they can be accelerated.  We consider the classic
example of a SN shock, although the discussion applies equally to
other shocks.  During a supernova explosion several solar masses
of material are ejected at a speed of $\sim 10^{4}$ km/s which is
much faster than the speed of sound in the interstellar medium
(ISM) which is $\sim$ 10 km/s.  A strong shock wave propagates
radially out as the ISM and its associated magnetic field piles
up in front of the supernova ejecta.  The velocity of the shock,
$V_{S}$, depends on the velocity of the ejecta, $V_{P}$, and on
the ratio of specific heats, $\gamma$ through the compression
ratio, $R$,
\begin{equation}
V_{S}/V_{P} \simeq R/(R-1).
\end{equation}
For SN shocks the SN will have ionized the surrounding gas which
will therefore be monatomic ($\gamma = 5/3$), and theory of shock
hydrodynamics shows that for $\gamma = 5/3$ a strong shock  will
have $R=4$.

\begin{figure}[htb]
\plotone{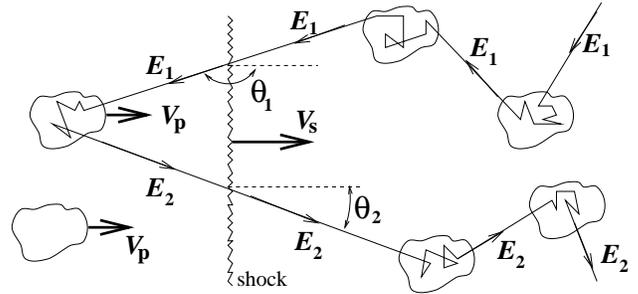}
\caption{Interaction of cosmic ray of energy $E_1$ with a shock moving with 
speed $V_s$.
\label{fig:fermi_acc_shock}}
\end{figure}

In order to work out the energy gain per shock crossing, we can
visualize magnetic irregularities on either side of the shock as
clouds of magnetized plasma of Fermi's original theory
(Fig.~\ref{fig:fermi_acc_shock}).  By considering the rate at
which cosmic rays cross the shock from downstream to upstream,
and upstream to downstream, one finds $\langle \cos \theta_{1}
\rangle = -2/3$ and $\langle \cos \theta_{2}^{\prime} \rangle =
2/3$, giving
\begin{equation}
{\langle \Delta E \rangle \over E}  \simeq {4 \over 3} \beta 
\simeq {4 \over 3} {V_{P} \over c} 
\simeq {4 \over 3} {(R-1) \over R} {V_{S} \over c}.
\end{equation}
Note this is 1st order in $\beta=V_{P}/c$ and is therefore more
efficient than Fermi's original theory.  This is because of the
converging flow -- whichever side of the shock you are on, if you
are moving with the plasma, the plasma on the other side of the
shock is approaching you at speed $V_p$.

\begin{figure}[htb]
\plotone{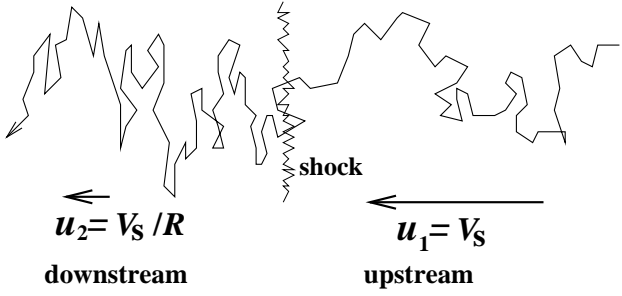}
\caption{Diffusion of cosmic rays from upstream to downstream
seen in the shock frame speed $V_s$.
\label{fig:up_downstream}}
\end{figure}

To obtain the energy spectrum we need to find the probability of
a cosmic ray encountering the shock once, twice, three times,
etc.  If we look at the diffusion of a cosmic ray as seen in the
rest frame of the shock (Fig.~\ref{fig:up_downstream}), there is
clearly a net flow of the energetic particle population in the
downstream direction.

The net flow rate downstream gives the rate at which cosmic rays are lost
downstream
\begin{equation}
r_{{\rm loss}} = n_{\rm CR} V_{S}/R \hspace{10mm} \rm m^{{-2}} s^{{-1}}
\label{eq:r_loss}
\end{equation}
since cosmic rays with number density $n_{\rm CR}$ at the shock
are advected downstream with speed $V_{S}/R$ (from right to left
in Fig.~\ref{fig:up_downstream}) and we have neglected
relativistic transformations of the rates because $V_S \ll c$.
 
Upstream of the shock, cosmic rays travelling at speed $v$ at
angle $\theta$ to the shock normal (as seen in the laboratory
frame) approach the shock with speed $(V_{S} + v \cos \theta)$ as
seen in the shock frame.  Clearly, to cross the shock, $\cos
\theta > -V_{S}/v$.  Then, assuming cosmic rays upstream are
isotropic, the rate at which they cross from upstream to
downstream is
\begin{eqnarray}
r_{\rm cross} &=&  n_{\rm CR} {1 \over 4 \pi} \int_{-V_S/v}^1 
(V_S +  v \cos \theta) 2 \pi d( \cos \theta) \nonumber \\
&\approx& n_{\rm CR} v/4 \hspace{5mm} \rm m^{-2} s^{-1}.
\label{eq:r_cross}
\end{eqnarray}

The probability of crossing the shock once and then escaping from
the shock (being lost downstream) is the ratio of these two
rates:
\begin{equation}
{\rm Prob.(escape)} = r_{\rm loss}/r_{\rm cross} \approx 4 V_{S}/Rv.
\end{equation}
The probability of returning to the shock
after crossing from upstream to downstream is
\begin{equation}
{\rm Prob.(return)} = 1 - {\rm Prob.(escape)},
\end{equation}
and so the probability of returning to the shock $k$ times and
also of crossing the shock at least $k$ times is
\begin{equation}
{\rm Prob.(cross} \ge k{\rm )} = [1 - {\rm Prob.(escape)}]^{k}.
\end{equation}
Hence, the energy after $k$ shock crossings is
\begin{equation}
E = E_{0} \left( 1 + {\Delta E \over E} \right)^{k}
\end{equation}
where $E_{0}$ is the initial energy.

To derive the spectrum, we note that the integral energy spectrum
(number of particles with energy greater than $E$) on
acceleration must be
\begin{equation}
Q(>E) \propto  [1 - {\rm Prob.(escape)}]^{k}
\end{equation}
where 
\begin{equation}
k = {\ln (E/E_{0}) \over \ln (1 + \Delta E/E)}.
\end{equation}
Hence,
\begin{equation}
\ln Q(>E) = A + {\ln (E/E_{0}) \over \ln (1 + \Delta E/E)} 
\ln [1 - {\rm Prob.(escape)}],
\end{equation}
where $A$ is a constant, and so
\begin{equation}
\ln Q(>E) = B - (\Gamma-1) \ln E
\end{equation}
where $B$ is a constant and
\begin{equation}
\Gamma = 1 - {\ln [1 - {\rm Prob.(escape)}] \over \ln (1 + \Delta
E/E)} \approx {R+2 \over R-1}.
\label{eq:gamma}
\end{equation}

Hence we arrive at the spectrum of cosmic rays on acceleration
\begin{equation}
Q(>E) \propto E^{{-(\Gamma-1)}} \hspace{1cm} \rm (integral \; form)
\end{equation}
\begin{equation}
Q(E) \propto E^{{-\Gamma}} \hspace{1cm} \rm (differential \; form).
\end{equation}
For $R=4$ we have the well-known $E^{-2}$ differential spectrum.
The observed cosmic ray spectrum is steepened by energy-dependent
escape of cosmic rays from the Galaxy.

\subsection{Effect of cosmic ray pressure}

For simplicity, we have neglected the effect of cosmic
ray pressure on the shock which can alter the shock profile.
The original method of treating this non-linear effect is the
two-fluid method, the two-fluids being plasma and cosmic rays, and
this is reviewed by Drury~(1983b)\markcite{Drury83b}.  The shock
profile, instead of being a step-function becomes smoothed, and
this affects the acceleration of low and high energy particles
differently.  Lower energy particles with short diffusion mean
free paths spanning only part of the shock profile
(i.e. effectively seeing a lower $R$) will have a steeper
spectrum, while high energy particles with longer diffusion mean
free paths spanning most of the shock profile (i.e. effectively
seeing a higher $R$) will have a flatter spectrum.  The net
result of this is a spectrum with upward curvature.  In addition,
as a result of the cosmic ray pressure far downstream providing
additional slowing of the plasma, the overall compression ratio
from far upstream to far downstream may be even higher than in
the test particle case giving an even flatter spectrum for the
highest energy particles (Ellison and Eichler
1984\markcite{EllisonEichler84}; see Baring
1997\markcite{Baring97} for a brief review and additional
references).

\subsection{Relativistic Shocks}

Shocks in jets of active galactic nuclei (AGN) and gamma ray
bursts (GRB) are likely to be relativistic, i.e. $u_1 > 0.1c$.
This will affect the acceleration in two ways: (a) the adiabatic
index of a relativistic gas is $\gamma=4/3$ giving $R=7$ for a
strong shock, and so one would expect from Eq.~\ref{eq:gamma} a
flatter spectrum with $\Gamma = 1.5$.  However, for relativistic
plasma motion with bulk velocities comparable to those of the
particles being accelerated, the approximations used to derive
Eqns.~\ref{eq:r_loss} and ~\ref{eq:r_cross} are no longer valid,
the escape probability being greater and the particles being
anisotropic, with the result that $\Gamma > 1.5$.  Detailed
studies have shown a trend in which shocks with larger $u_1$
generally have lower $\Gamma$~(Kirk and Schneider 1987, Ellison
and Jones
1990)\markcite{KirkSchneider87,EllisonJones90}. However, the
spectral index is very sensitive to the pitch angle (see Baring
1997\markcite{Baring97} for additional references).

\section{Shock Acceleration Rate}

Here we again neglect effects of cosmic ray pressure and
consider only a non-relativistic shock.  The acceleration rate is defined by
\begin{equation}
r_{\rm acc} \equiv {1 \over E} \left. {dE \over dt} \right|_{\rm
acc} = {(\langle \Delta E \rangle /E) \over t_{\rm cycle}}
\approx {4 \over 3} {(R-1) \over R} {V_{S} \over c} t_{\rm
cycle}^{-1}
\end{equation}
where $t_{\rm cycle}$ is the time for one complete cycle, i.e.
from crossing the shock from upstream to downstream, diffusing
back towards the shock and crossing from downstream to upstream,
and finally returning to the shock. 

The rate of loss of accelerated particles downstream is the probability
of escape per shock crossing divided by the cycle time
\begin{equation}
r_{\rm esc} = {{\rm Prob.(escape)} \over t_{\rm cycle}} \approx
{4 \over R} {V_{S} \over c} t_{\rm cycle}^{-1}
\end{equation}

We see immediately that the ratio of the escape rate to the
acceleration rate depends on the compression ratio
\begin{equation}
{r_{\rm esc} \over r_{\rm acc}} \approx {3 \over R-1}
\end{equation}
and for a strong shock ($R=4$) the two rates are equal.  As we
shall see later, a consequence of this is that the asymptotic
spectrum of particles accelerated by a strong shock is the well-known
$E^{-2}$ power-law.

We shall discuss these processes in the shock frame (see
Fig.~\ref{fig:up_downstream}) and consider first particles
crossing the shock from upstream to downstream and diffusing back
to the shock, i.e.\ we shall work out the average time spent
downstream.  Since we are considering non-relativistic shocks,
the time scales are approximately the same in the upstream and
downstream plasma frames, and so in this section I shall drop the
use of subscripts indicating the frame of reference.

Diffusion takes place in the presence of advection at speed $u_2$
in the downstream direction.  The typical distance a particle
diffuses in time t is $\sqrt{k_2t}$ where $k_2$ is the diffusion
coefficient in the downstream region.  The distance advected in
this time is simply $u_2t$.  If $\sqrt{k_2t} \gg u_2t$ the
particle has a very high probability of returning to the shock,
and if $\sqrt{k_2t} \ll u_2t$ the particle has a very high
probability of never returning to the shock (i.e. it has
effectively escaped downstream).  So, we set $\sqrt{k_2t} = u_2t$
to define a distance $k_2/u_2$ downstream of the shock which is
effectively a boundary between the region closer to the shock
where the particles will usually return to the shock and the
region farther from the shock in which the particles will usually
be advected downstream never to return.  There are $n_{\rm CR}
k_2/u_2$ particles per unit area of shock between the shock and
this boundary.  Dividing this by $r_{\rm cross}$ we obtain the
average time spent downstream before returning to the shock
\begin{equation}
t_2 \approx {4 \over c} {k_2 \over u_2}.
\end{equation}

Consider next the other half of the cycle after the particle has
crossed the shock from downstream to upstream until it returns to
the shock.  In this case we can define a boundary at a distance
$k_1/u_1$ upstream of the shock such that nearly all particles
upstream of this boundary have never encountered the shock, and
nearly all the particles between this boundary and the shock have
diffused there from the shock.  Then dividing the number of
particles per unit area of shock between the shock and this
boundary, $n_{\rm CR} k_1/u_1$, by $r_{\rm cross}$ we obtain the
average time spent upstream before returning to the shock
\begin{equation}
t_1 \approx {4 \over c} {k_1 \over u_1},
\end{equation}
and hence the cycle time
\begin{equation}
t_{\rm cycle} \approx {4 \over c} \left( {k_1 \over u_1} + {k_2
\over u_2} \right).
\end{equation}

The acceleration rate is then given by
\begin{equation}
r_{\rm acc} \approx {(R-1)u_1 \over 3R} \left( {k_1 \over u_1} + {k_2
\over u_2} \right)^{-1}.
\end{equation}
Assuming that the diffusion coefficients upstream and downstream
have the same power-law dependence on energy, e.g.,
\begin{equation}
k_1 \propto k_2 \propto E^{\delta},
\end{equation}
then the acceleration rate also has a power-law dependence
\begin{equation}
r_{\rm acc}\propto E^{-\delta}.
\end{equation}
Note that more correctly, the diffusion coefficient will be a
function of magnetic rigidity, $\rho$, which, for
ultra-relativistic particles considered in this paper, is
approximately equal to $E/Ze$ where $Ze$ is the charge.  However,
here we are mainly concerned with singly charged particles and
shall work in terms of $E$ rather than rigidity.

\subsection{Maximum acceleration rate}

We next consider the diffusion for the cases of parallel,
oblique, and perpendicular shocks, and estimate the maximum
acceleration rate for these cases.  The diffusion coefficients
required, $k_1$ and $k_2$, are the coefficients for diffusion
parallel to the shock normal.  The diffusion coefficient along
the magnetic field direction is some factor $\eta$ times the
minimum diffusion coefficient, known as the Bohm diffusion
coefficient,
\begin{equation}
k_\parallel = \eta {1 \over 3} r_g c
\end{equation}
where $r_g$ is the gyroradius, and $\eta > 1$.

Parallel shocks are defined such that the shock normal is
parallel to the magnetic field ($\vec{B} || \vec{u_1}$).  In this
case, making the approximation that $k_1 = k_2 = k_\parallel$ and
$B_1 = B_2$ one obtains
\begin{equation}
t_{\rm acc}^\parallel \approx {20 \over 3} {\eta E \over e B_1 u_1^2}.
\end{equation}
For a shock speed of $u_1 = 0.1 c$ and $\eta=10$ one obtains an
acceleration rate (in SI units) of
\begin{equation}
\left. {dE \over dt} \right|_{\rm acc} \approx 1.5 \times 10^{-4}
e c^2 B.
\end{equation}

For the oblique case, the angle between the magnetic field
direction and the shock normal is different in the upstream and
downstream regions, and the direction of the plasma flow also
changes at the shock.  The diffusion coefficient in the direction
at angle $\theta$ to the magnetic field direction is given by
\begin{equation}
k = k_\parallel \cos^2 \theta + k_\perp \sin^2 \theta
\end{equation}
where $k_\perp$ is the diffusion coefficient perpendicular to the
magnetic field.  Jokipii (1987)\markcite{Jokipii87} shows that
\begin{equation}
k_\perp \approx {k_\parallel \over 1 + \eta^2}
\end{equation}
provided that $\eta$ is not too large (values in the range up to
10 appear appropriate), and that acceleration at perpendicular
shocks can be much faster than for the parallel case.  For
$k_{xx} = k_\perp$ and $B_2 \approx 4 B_1$ and one obtains
\begin{equation}
t_{\rm acc}^\perp \approx {8 \over 3} {E \over \eta e B_1 u_1^2}.
\end{equation}
For a shock speed of $u_1 = 0.1 c$ and $\eta=10$ one obtains an
acceleration rate (in SI units) of
\begin{equation}
\left. {dE \over dt} \right|_{\rm acc} \approx 0.04 e c^2 B.
\end{equation}

Supernova shocks remain strong enough to continue accelerating
cosmic rays for about 1000 years.  The rate at which cosmic rays
are accelerated is inversely proportional to the diffusion
coefficient (faster diffusion means less time near the shock).
For the maximum feasible acceleration rate, a typical
interstellar magnetic field, and 1000 years for acceleration,
energies of $10^{{14}} \times Z$ eV are possible ($Z$ is atomic
number) at parallel shocks and $10^{{16}} \times Z$ eV at
perpendicular shocks.

In the case of acceleration by relativistic shocks (e.g. Bednarz
\& Ostrowski~1996, 1998, and Bednarz 1998, and
Ostrowski~1998)\markcite{BednarzOstrowski96,BednarzOstrowski98,
Bednarz98,Ostrowski98}, the acceleration time depends strongly on
$u_1$, $k_\perp/k_\parallel$, and $\theta$ and can be as low as
$\sim (1$---10$)r_g/c$.

%%%%%%%%%%%%%%%%%%%%%%%%%%%%%%%%%%%%%%%%%%%%%%%%%%%%%%%%%%%%%%%%%%%%%%%%%%%%%%%

%%%%%%%%%%%%%%%%%%%%%%%%%%%%%%%%%%%%%%%%%%%%%%%%%%%%%%%%%%%%%%%%%%%%%%%%%%%%%%%

\section{Maximum Energies}

Protons and nuclei can be accelerated to much higher energies
than electrons for a given magnetic environment.  For stochastic
particle acceleration by electric fields induced by motion of
magnetic fields $B$, the rate of energy gain by relativistic
particles of charge $Ze$ can be written (in SI units) $dE/dt =
\xi Ze c^2 B$ as in Eq.~1, where $\xi < 1$ and depends on the
acceleration mechanism; a value of $\xi =0.04$ might be achieved
by first order Fermi acceleration at a perpendicular shock with
shock speed $\sim 0.1 c$.

\subsection{Limits from Synchrotron Losses}

The rate of energy loss by synchrotron radiation of a particle of mass
$Am_p$, charge $Ze$, and energy $\gamma m c^2$ is
\begin{equation}
- \left. {dE \over dt} \right|_{\rm syn} = {4 \over 3} \sigma_T
\left( {Z^2 m_e \over Am_p} \right)^2 {B^2 \over 2 \mu_0} \gamma^2 c.
\end{equation}
Equating the rate of energy gain with the rate of energy loss by
synchrotron radiation places one limit on the maximum energy achievable
by electrons, protons and nuclei:
\begin{eqnarray}
E_e^{\rm max} &=& 6.0 \times 10^{2} \xi^{1/2}
\left( {B \over {\rm 1 \; T}} \right)^{-1/2} \hspace{5mm} {\rm GeV}, \\
E_p^{\rm max} &=& 2.0 \times 10^{9} \xi^{1/2}
\left( {B \over {\rm 1 \; T}} \right)^{-1/2} \hspace{5mm} {\rm GeV}, \\
E_{Z,A}^{\rm max} &=& 2.0 \times 10^{9} \xi^{1/2} {A^2 \over Z^{3/2}}
\left( {B \over {\rm 1 \; T}} \right)^{-1/2} \, {\rm GeV}.
\end{eqnarray}
The cut-off in proton spectrum may be caused by synchrotron
radiation, and Totani (1998)\markcite{Totani98} has found
that synchrotron radiation by highest energy protons could account
for much of the observed cosmic $\gamma$-ray background.  The
maximum energies of protons and iron nuclei allowed by
synchrotron radiation losses are shown in
Figs.~\ref{fig:pgpiee3k_emax} and \ref{fig:fe_emax} respectively,
and are plotted against magnetic field for three values of $\xi$.
The straight lines are for: the maximum possible acceleration rate
$\xi=1$ (dashed), plausible acceleration at perpendicular shock
$\xi=0.04$ (solid), and plausible acceleration at parallel shock
$\xi=1.5 \times 10^{-4}$ (dot-dash).

Other limits on the maximum energy are placed by the dimensions of
the acceleration region and the time available for acceleration.  These 
limits were obtained and discussed in some detail by 
Biermann and Strittmatter (1987)\markcite{BS87}.

\subsection{Limits from Interactions with Radiation}

\begin{figure}[htb]
\plotone{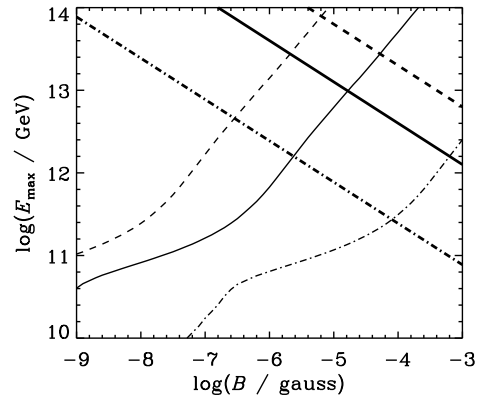}
\caption{Maximum proton energy as a function of magnetic field.
Straight lines give the limit from synchrotron loss, curved lines
give the limit from pion photoproduction.  $E_{\rm max}$ is given for
$\xi=1$ (dashed), $\xi=0.04$ (solid), $\xi=1.5 \times 10^{-4}$
(dot-dash).
\label{fig:pgpiee3k_emax}}
\end{figure}

Equating the total energy loss rate for proton--photon
interactions (i.e. the sum of pion production and Bethe-Heitler
pair production) in Fig.~\ref{fig:pgpiee3k_xloss} to the rate of
energy gain by acceleration gives the maximum proton energy in
the absence of other loss processes.  This is shown in
Fig.~\ref{fig:pgpiee3k_emax} as a function of magnetic field which
determines the rate of energy gain through Eq.~\ref{eq:max_acc}.
The result is shown by the curves for the maximum possible
acceleration rate $\xi=1$ (dashed), plausible acceleration at
perpendicular shock $\xi=0.04$ (solid), and plausible
acceleration at parallel shock $\xi=1.5 \times 10^{-4}$
(dot-dash).  Also shown is the maximum energy determined by
synchrotron losses (thick lines) for the three cases.  As can be
seen, for a perpendicular shock it is possible to accelerate
protons to $\sim 10^{13}$ GeV in a $\sim 10^{-5}$ G
field.

\begin{figure}[htb]
\plotone{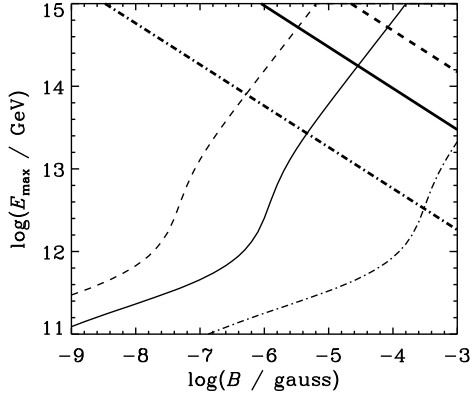}
\caption{Maximum iron nucleus energy as a function of magnetic field.
Straight lines give the limit from synchrotron loss, curved lines
give the limit from photodisintegration.  $E_{\rm max}$ is given for
$\xi=1$ (dashed), $\xi=0.04$ (solid), $\xi=1.5 \times 10^{-4}$
(dot-dash).
\label{fig:fe_emax}}
\end{figure}

The effective loss distance given in Fig.~\ref{fig:fepiee_xloss}
is used together with the acceleration rate for iron nuclei to
obtain the maximum energy as a function of magnetic field.  This
is shown in Fig.~\ref{fig:fe_emax} which is analogous to
Fig.~\ref{fig:pgpiee3k_emax} for protons.  We see that for a
perpendicular shock it is possible to accelerate iron nuclei to
$\sim 2 \times 10^{14}$ GeV in a $\sim 3 \times 10^{-5}$ G field.
While this is higher than for protons, iron nuclei are likely to
get photodisintegrated into nucleons of maximum energy $\sim 4
\times 10^{12}$ GeV, and so there is not much to be gained
unless the source is nearby.  Of course, potential acceleration
sites need to have the appropriate combination of size (much
larger than the gyroradius at the maximum energy), magnetic
field, and shock velocity (or other relevant velocity), and these
criteria have been discussed in detail by Hillas
(1984)\markcite{Hil84}.

%%%%%%%%%%%%%%%%%%%%%%%%%%%%%%%%%%%%%%%%%%%%%%%%%%%%%%%%%%%%%%%%%%%%%%%%%%%%%%%

\section{Spectral Shape near Maximum Energy}

To determine the spectral shape near maximum energy we use the
leaky-box acceleration model (Szabo and Protheroe
1994)\markcite{SzaboProtheroe94} which may be considered as
follows.  A particle of energy $E_0$ is injected into the leaky
box.  While inside the box, the particle's energy changes at a
rate $dE/dt = E r_{\rm acc}(E)$ and that in any short time
interval $\Delta t$ the particle has a probability of escaping
from the box given by $\Delta t r_{\rm esc}(E)$.  The energy
spectrum of particles escaping from the box then approximates the
spectrum of shock accelerated particles.

Let us consider first the case of no energy losses, interactions,
or losses due to any other process.  $N_0$ particles of energy
$E_0$ are injected at time $t=0$, and we assume the following
acceleration and escape rates:
\begin{equation}
r_{\rm acc} = a E^{-\delta},
\end{equation}
\begin{equation}
r_{\rm esc} = c E^{-\delta}.
\end{equation}
The energy at time $t$ is then obtained simply by integrating
\begin{equation}
dE/dt = aE^{(1-\delta)},
\label{eq:accrate}
\end{equation}
giving
\begin{equation}
E(t) = (E_0^\delta + \delta a t)^{1/\delta}.
\label{eq:E(t)}
\end{equation}
The number of particles remaining inside
the accelerator at time $t$ after injection is obtained by solving
\begin{equation}
dN/dt = - N(t) c [E(t)]^{-\delta}.
\end{equation}
Using Eq.~\ref{eq:E(t)} and integrating, one has  
\begin{equation}
\int_{N_0}^{N(t)} N^{-1} dN = - c \int_0^t (E_0^\delta + \delta a
t)^{-1} dt,
\end{equation}
giving
\begin{equation}
N(t) = N_0 [E(t) / E_0]^{-c/a}.
\end{equation}
Since $N_0 - N(t)$ particles have escaped from the accelerator
before time $t$, and therefore have energies between $E_0$ and
$E(t)$, the differential energy spectrum of particles which
have escaped from the accelerator is simply given by
\begin{eqnarray}
dN/dE & = & N_0 (\Gamma - 1) (E_0)^{- 1} (E/E_0)^{- \Gamma}, 
\label{eq:spec_nocut}
\end{eqnarray}
for $(E>E_0)$ where $\Gamma = (1+c/a)$ is the differential spectral index.  We
note that for $r_{\rm esc}(E) = r_{\rm acc}(E)$ one obtains the
standard result for acceleration at strong shocks $\Gamma = 2$.

\subsection{Cut-off due to finite acceleration volume, etc.}

Even in the absence of energy losses, acceleration usually ceases
at some energy due to the finite size of the acceleration volume
(e.g. when the gyroradius becomes comparable to the characteristic
size of the shock), or as a result of some other process.  We
approximate the effect of this by introducing a constant term to
the expression for the escape rate:
\begin{equation}
r_{\rm esc} = c E^{-\delta} + c E_{\rm max}^{-\delta}.
\label{eq:escrate}
\end{equation}
where $E_{\rm max}$ is defined by the above equation and will
be close to the energy at which the spectrum steepens due
to the constant escape term.  We shall refer to $E_{\rm max}$ 
as the ``maximum energy'' even though some particles will be
accelerated to energies above this.

\begin{figure}[htb]
\plotone{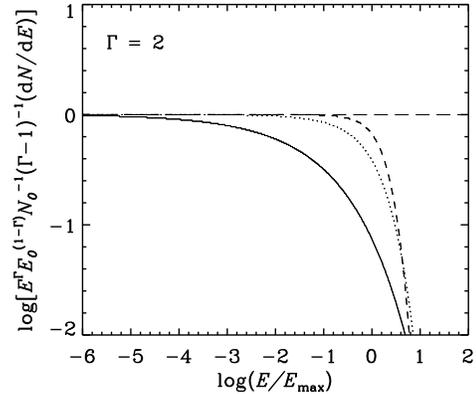}
\caption{Differential energy spectrum for $\Gamma=2$ and
$\delta=1/3$ (solid curve), 2/3 (dotted curve) and 1 (dashed
curve). (From Protheroe \& Stanev 1998).
\label{leakage_cutoff_comp}}
\end{figure}

Following the same procedure as for the case of a purely
power-law dependence of the escape rate, we obtain the
differential energy spectrum of particles $(E>E_0)$ escaping from the
accelerator,
\begin{eqnarray}
{dN \over dE} \! &=& \! N_0 (\Gamma - 1) (E_0)^{- 1} (E/E_0)^{-
\Gamma}[ 1 + ( {E / E_{\rm max}} )^\delta ] \nonumber \\ && \hspace{-5mm}
\times \exp \left\{ - {\Gamma - 1 \over \delta} \left[ \left( {E
\over E_{\rm max}} \right)^\delta - \left( {E_0 \over E_{\rm
max}} \right)^\delta \right] \right\}
\end{eqnarray}
for $\delta>0$.  For $\delta = 0$ we note that from
Eq.~\ref{eq:escrate} $r_{\rm esc}=(2c)$ at all energies.
Thus the situation is equivalent to the case of $E_{\rm max} \to
\infty$ and the spectrum is given by Eq.~\ref{eq:spec_nocut}
provided we replace $c$ with $2c$, i.e. we must replace $\Gamma$
with $\Gamma' =(2 \Gamma-1)$.  
We compare in Fig~\ref{leakage_cutoff_comp} the spectra for
$\Gamma=2$ and $\delta$ ranging from $1/3$ to $1$, and note that
the energy dependence of the diffusion coefficient has a profound
influence on the shape of the cut-off.

\vspace{5mm}

\vspace{15mm}
Spectra such as those presented in Fig.~\ref{leakage_cutoff_comp}
may be used to model the source spectra of high energy cosmic ray nuclei
of various species if one replaces energy with magnetic
rigidity, $\rho = pc/Ze \approx E/Ze$, where $Z$ is the atomic number:
\begin{eqnarray}
{dN \over d\rho} \! &=& \!N_0 (\Gamma - 1) (\rho_0)^{- 1}
(\rho/\rho_0)^{- \Gamma} [ 1 + ( {\rho / \rho_{\rm max}} )^\delta
] \nonumber \\ && \hspace{-5mm} \times \exp \left\{ - {\Gamma - 1
\over \delta} \left[ \left( {\rho \over \rho_{\rm max}}
\right)^\delta - \left( {\rho_0 \over \rho_{\rm max}}
\right)^\delta \right] \right\}
\end{eqnarray}
for $\rho>\rho_0$.

\subsection{Cut-off due to $E^2$ energy losses}

$E^2$ energy losses of protons or nuclei at extremely high energies result
from synchrotron radiation.  The introduction of continuous
energy losses into the problem is, in principle, straightforward
and accomplished simply by modifying Eq.~\ref{eq:accrate},
\begin{equation}
dE/dt = a E^{(1-\delta)} -b E^2.
\label{eq:acclossrate}
\end{equation}
Setting $dE/dt = 0$ we obtain the cut-off energy
\begin{equation}
E_{\rm cut}= (a/b)^{1/(1+\delta)}.
\end{equation}
For this case, however, the problem is easier to solve using the
Green's function approach adopted by Stecker~(1971)\markcite{Stecker71}
when considering the ambient spectrum of cosmic ray electrons and
galactic $\gamma$-rays.  Using the appropriate Green's function one
can obtain the steady-state spectrum of particles inside the
leaky-box accelerator, and multiplying this by the escape rate
one obtains the spectrum of particles leaving the accelerator,
\begin{eqnarray}
{d N \over dE} = {\frac{c\, {E^{-\delta}} + {E_{\rm max}^{-\delta}} }
       {a\,{E^{1 - \delta}}- b\,{E^2}}} \exp [ - I(E)]
\end{eqnarray}
where
\begin{eqnarray}
I(E) & \equiv & \int_{E_0}^E {\frac{c\, {E^{-\delta}} + {E_{\rm
max}^{-\delta}} } {a\,{E^{1 - \delta}}- b\,{E^2}}}\,dE\\ &=&
\left[ \left( {c \over a\,\delta} \right) \left( {E \over E_{\rm
max}} \right)^\delta \right. \nonumber \\ && \times
\left. {_2F_1} \! \left( {\frac{\delta}{1 + \delta}},1, 1 +
{\frac{\delta}{1 + \delta}},{\frac{b\,{E^{1 + \delta}}}{a}}
\right) \right. \nonumber \\ && \left. - \;\frac{c}{a( 1 +
\delta) }\ln \! \left( b - {a \over E^{1 + \delta}} \right)
\right]_{E_0}^E ,
\end{eqnarray}
and ${_2F_1}$ is the hypergeometric function.

\begin{figure}[htb]
\plotone{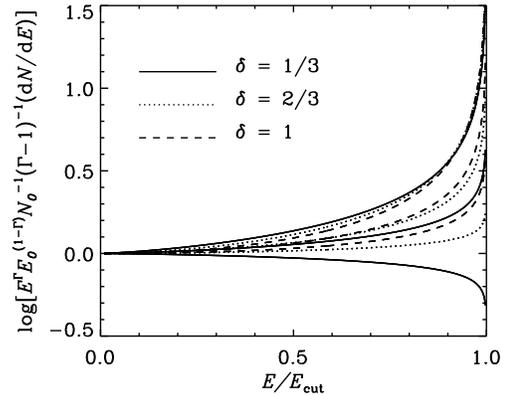}
\caption{Differential energy spectra for $E_0 \ll E_{\rm cut} \ll
E_{\rm max}$ for $\delta=1/3$ (solid curves), 2/3 (dotted curves)
and 1 (dashed curves), and $\Gamma=1.5$ (upper curves), 2.0
(middle curves) and 2.5 (lower curves). (From Protheroe \& Stanev 1998).
\label{synch_pileup}}
\end{figure}

The result depends on the parameters
$\delta$, $\Gamma$, $E_0$, $E_{\rm cut}$, $E_{\rm max}$.
As a result of the energy loss by particles near the maximum
energy a pile-up in the spectrum may be produced just below
$E_{\rm cut}$.  The size of the pile-up will be determined by the
relative importance of $r_{\rm acc}$ and $r_{\rm esc}$ at
energies just below $E_{\rm cut}$, and so should depend only on
$\Gamma$ and $\delta$ provided $E_0 \ll E_{\rm cut} \ll E_{\rm
max}$.  In this case, the shape of the spectrum is given by
\begin{eqnarray}
{d N \over dE} &=& N_0 (\Gamma - 1) (E_0)^{- 1} \left( {E \over
E_0} \right)^{- \Gamma} \nonumber \\ && \times \left[ 1 - \left(
{E \over E_{\rm cut}} \right)^{(1 + \delta)} \right]^{(\Gamma -2
-\delta)/(1 + \delta)}.
\end{eqnarray}
We compare in Fig.~\ref{synch_pileup} the spectra for this case
for $\delta=1/3$, 2/3 and 1, and for $\Gamma = 1.5$, 2, and 2.5,
and note that the pile-ups are higher for flatter spectra, and
that for steep spectra the pile-up may be absent or the spectrum
may steepen before the cut-off if $\delta$ is small.  The effect
of synchrotron losses on multiple diffusive shock acceleration
have been considered by Melrose and Crouch (1997)
\markcite{MelroseCrouch}, and similar effects were found.

\subsection{Cut-off due to pion photoproduction}

Here I discuss how one can use the Monte Carlo method to
investigate the shape of the cut-off or pile-up which results
when the nominal cut-off energy is determined by interactions
rather than continuous energy losses.  This technique was used
by Protheroe and Stanev (1998) to investigate cut-offs in electron spectra
due to inverse Compton scattering in the Klein-Nishina regime,
and by Szabo and Protheroe(1994) to investigate cut-offs in the proton
spectrum due to photoproduction in a radiation field.

The technique uses the leaky-box acceleration model described
above.  We describe here the case considered by Szabo and
Protheroe: $\Gamma=2, \delta=1$.  Particles of energy $E_0$ are
injected and accelerated at a constant rate $a={\rm d}E/{\rm
d}t$.  The time scale for escape from the leaky box is equal to
the acceleration time scale, $t_{\rm esc}=t_{\rm acc}=E/a$, and
so the probability of reaching an energy $E$ without escaping is
$P_{\rm surv}(E)=E_0/E$.  Such a model produces an $E^{-2}$
spectrum at energies above $E_0$ and may be easily adapted to
calculations in which other processes take place in addition to
acceleration, and are simulated by the Monte Carlo method.

To calculate the spectrum of particles produced during
acceleration per injected particle, we use a method of weights.
If a particle is injected with energy $E_0$, then the ``weight''
of the particle by the time it reaches an energy $E_1$ is set to
the probability of it not having escaped, $W_1=E_0/E_1$.  If the
particle interacts, and has energy $E_1^{\prime}$ after the
interaction, etc., then the weight of the particle at the $n$th
interaction is given by
\begin{equation}
	W_n=\frac{E_0}{E_1}\frac{E_1^{\prime}}{E_2}\cdots\frac
	{E_{n-1}^{\prime}}{E_{n}}.
\end{equation}
For the case in which $E_{\rm max}$ is determined by
proton-photon collisions, the following procedure was adopted.
(i) Inject a particle with energy $E_0$ into the accelerator and
after $(n-1)$ interactions (assuming it has survived catastrophic
losses) it will have energy $E_{n-1}^{\prime}$ and weight
$W_{n-1}$; (ii) Sample the $n$th path length for two competing
interactions, the competing interaction with the shortest path
length being assumed to be the one that occurs.

Before the interaction, the proton has energy $E_n$ and the
interaction is modelled by the Monte Carlo method.
The energies of the produced particles are binned in energy with
the appropriate weight, in this case $W_n$.  The `fraction' of
the particle which does not interact, in this case $(1-W_n)$, is
assumed to escape from the accelerator and have an $E^{-2}$
energy distribution over the energy range $E_{n-1}^{\prime}$ to
$E_n$.  This contribution is then added to the escaping proton
spectrum.  The process is repeated until the particle suffers a
catastrophic loss in the form of a pion photoproduction reaction
in which a neutron is produced.

\begin{figure}[htb]
\plotone{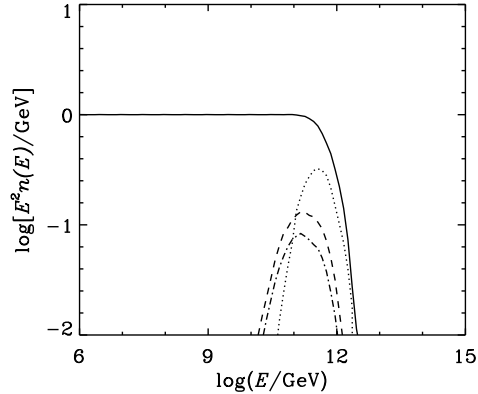}
\caption{The spectrum of particles produced during acceleration
(multiplied by $E^2$) per proton injected into the accelerator:
protons (full curve), neutrons (dotted curve), charged pions
(dashed curve) and neutral pions. Results are shown for $E_{\rm
cut}= 2 \times 10^{12}$ GeV due to photoproduction on the cosmic
microwave background.  (Adapted from Fig.~7 of Szabo and
Protheroe 1994).
\label{fig:AccSpec98}}
\end{figure}

In this way one can calculate the spectra of protons, neutrons,
pions (neutral and charged) and electrons (including positrons)
produced during acceleration.  Results for $E_{\rm cut}= 2 \times
10^{12}$ GeV due to photoproduction on the cosmic microwave
background radiation are shown in Figure~\ref{fig:AccSpec98}.

%%%%%%%%%%%%%%%%%%%%%%%%%%%%%%%%%%%%%%%%%%%%%%%%%%%%%%%%%%%%%%%%%%%%%%%%%%%%%%%
 
\section{Cascading During Propagation}

As well as particles being produced during the acceleration
process as a result of interactions, during propagation to Earth
cascading occurs and the accompanying fluxes of $\gamma$-rays and
neutrinos must not exceed the observed flux or flux limits.
Waxman and Bahcall (1998) estimate the maximum allowed neutrino
flux to be very low and model independent in the important TeV
to PeV range.  However, Mannheim et
al. (1998)\markcite{Mannheim98} obtain a limit in this energy
range which is orders of magnitude higher.  By measuring the
accompanying fluxes, we may well provide additional clues to the
nature and origin of the highest energy cosmic rays.  Hence it is
important to calculate these fluxes resulting from cascading.

There are several cascade processes which are important for UHE
cosmic rays propagating over large distances through a radiation
field: protons interact with photons resulting in pion production
and pair production; electrons interact via inverse-Compton
scattering and triplet pair production, and emit synchrotron
radiation in the intergalactic magnetic field; $\gamma$-rays
interact by pair production.  Energy losses due to cosmological
redshifting of high energy particles and $\gamma$-rays can also
be important, and the cosmological redshifting of the background
radiation fields means that energy thresholds and interaction
lengths for the above processes also change with epoch (see
e.g. Protheroe et
al. (1995)\markcite{ProtheroeStanevBerezinsky}).

\begin{figure}[htb]
\plotone{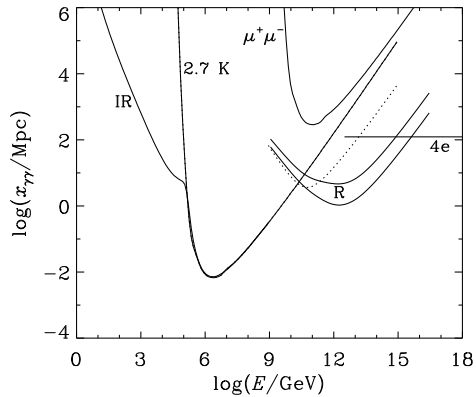}
\caption{The mean interaction length for pair production for
$\gamma$-rays in the Radio Background calculated in the present
work (solid curves labelled R: upper curve -- no evolution of
normal galaxies; lower curve -- pure luminosity evolution of
normal galaxies) and in the radio background of Clark (1970)
\protect \markcite{Cla70} (dotted line).  Also shown are the mean
interaction length for pair production in the microwave
background (2.7K), the infrared and optical background (IR), and
muon pair production ($\mu^+\mu^-$) and double pair production
(4e) in the microwave background (Protheroe and Johnson 1995).
\protect \markcite{ProtheroeJohnson95}  (From Protheroe and
Biermann 1996\protect \markcite{ProtheroeBiermann96a}).
\label{fig:ggee_radio}}
\end{figure}

The energy density of the extragalactic background radiation is
dominated by that from the cosmic microwave background at a
temperature of 2.73 K.  Other components of the extragalactic
background radiation are discussed in the review of Ressel and
Turner (1990)\markcite{Res90}.  The extragalactic radiation
fields important for cascades initiated by UHE cosmic rays
include the cosmic microwave background, the radio background and
the infrared--optical background.  The radio background was
measured over twenty years ago (Bridle 1967, Clarke et
al. 1970)\markcite{Bri67,Cla70}, but the fraction of this radio
background which is truly extragalactic, and not contamination
from our own Galaxy, is still debatable.  Berezinsky
(1969)\markcite{Ber69} was first to calculate the mean free path
on the radio background.  More recently Protheroe and Biermann
(1996)\markcite{ProtheroeBiermann96a} have made a new calculation
of the extragalactic radio background radiation down to kHz
frequencies.  The main contribution to the background is from
normal galaxies and is uncertain due to uncertainties in their
evolution.  The mean free path of photons in this radiation field
as well as in the microwave and infrared backgrounds is shown in
Fig.~\ref{fig:ggee_radio}.  Also shown is the mean interaction
length for muon pair-production which is negligible in comparison
with interactions with pair production on the radio background
and double pair production on the microwave background.

\begin{figure}[htb]
\plotone{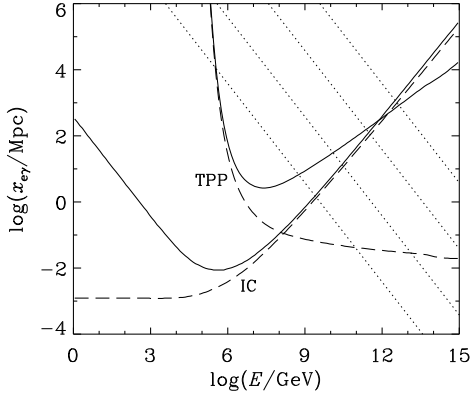}
\caption{The mean interaction length (dashed line) and
energy-loss distance (solid line), $E/(dE/dx)$, for
electron-photon triplet pair production (TPP) and inverse-Compton
scattering (IC) in the microwave background.  The energy-loss
distance for synchrotron radiation is also shown (dotted lines)
for intergalactic magnetic fields of $10^{-9}$ (bottom),
$10^{-10}$, $10^{-11}$, and $10^{-12}$ gauss (top).  (From
Protheroe and Johnson 1995.) \protect
\markcite{ProtheroeJohnson95}
\label{fig:e3kmfp}}
\end{figure}

Inverse Compton interactions of high energy electrons and triplet
pair production can be modelled by the Monte Carlo technique
(e.g. Protheroe 1986, Protheroe 1990, Protheroe et al. 1992,
Mastichiadis et al. 1994)\markcite{Pro86,Pro90,Pro92,Mas94}, and
the mean interaction lengths and energy-loss distances for these
processes are given in Fig. \ref{fig:e3kmfp}.  Synchrotron losses
must also be included in calculations and the energy-loss
distance has been added to Fig. \ref{fig:e3kmfp} for various
magnetic fields.

%%%%%%%%%%%%%%%%%%%%%%%%%%%%%%%%%%%%%%%%%%%%%%%%%%%%%%%%%%%%%%%%%%%%%%%%%%%%%%%
\subsection{Practical Aspects of the Cascade}

%%%%%%%%%%%%%%%%%%%%%%%%%%%%%%%%%%%%%%%%%%%%%%%%%%%%%%%%%%%%%%%%%%%%%%%%%%%%%%%
Where possible, to take account of the exact energy dependences
of cross-sections, one can use the Monte Carlo method.  However,
direct application of Monte Carlo techniques to cascades
dominated by the physical processes described above over
cosmological distances takes excessive computing time.  Another
approach based on the matrix multiplication method has been
described by Protheroe (1986) \markcite{Pro86} and developed in
later papers (Protheroe and Stanev 1993, Protheroe and Johnson
1995)\markcite{Pro93,ProtheroeJohnson95}.  A Monte Carlo program
is used to calculate the yields of secondary particles due to
interactions with radiation, and spectra of produced pions are
decayed (e.g. using routines in SIBYLL (Fletcher et
al. 1994)\markcite{Fle94}) to give yields of $\gamma$-rays,
electrons and neutrinos.  For the pion photoproduction
interactions a new program called SOPHIA is available (M\"{u}cke
et al. 1998\markcite{Muecke98}).  The yields are then used to
build up transfer matrices which describe the change in the
spectra of particles produced after propagating through the
radiation fields for a distance $\delta x$.  Manipulation of the
transfer matrices as described below enables one to calculate the
spectra of particles resulting from propagation over arbitrarily
large distances.

%%%%%%%%%%%%%%%%%%%%%%%%%%%%%%%%%%%%%%%%%%%%%%%%%%%%%%%%%%%%%%%%%%%%%%%%%%%%%%%
\subsection{Matrix Method}
%%%%%%%%%%%%%%%%%%%%%%%%%%%%%%%%%%%%%%%%%%%%%%%%%%%%%%%%%%%%%%%%%%%%%%%%%%%%%%%
In the work of Protheroe and Johnson
(1995)\markcite{ProtheroeJohnson95}, fixed logarithmic energy
bins were used, and the energy spectra of particles of type
$\alpha$ ($\alpha =
\gamma,e,p,n,\nu_e,\bar{\nu}_e,\nu_\mu,\bar{\nu}_\mu$) at
distance $x$ in the cascade are represented by vectors
$F_{j}^{\alpha}(x)$ which give the total number of particles of
type $\alpha$ in the $j$th energy bin at distance $x$.  Transfer
matrices, T$_{ij}^{\alpha\beta}(\delta x)$, give the number of
particles of type $\beta$ in the bin $j$ which result at a
distance $\delta x$ after a particle of type $\alpha$ and energy
in bin $i$ initiates a cascade.  Then, given the spectra of
particles at distance $x$ one can obtain the spectra at distance
$(x + \delta x)$
\begin{equation}
F_{j}^{\beta}(x +\delta x) = \sum_{\alpha}
        \sum_{i=j}^{180} {\rm T}_{ij}^{\alpha \beta}
        (\delta x)F_{i}^{\alpha}(x)
\end{equation}
where $F_{i}^{\alpha}(x)$ are the input spectra
(number in the $i$th energy bin) of species $\alpha$.

We could also write this as
\begin{equation}
[{\rm F}(x +\delta x)] = [{\rm T}(\delta x)][{\rm F}(x)]
\end{equation}
where 
\begin{equation}
\rm
[F] = \left[ \begin{array}{c} F^{\gamma} \\ F^{e} \\ F^{p} 
\\ \vdots  \end{array} \right] , \hspace{5mm}
[T] = \left[ \begin{array}{cccc} 
{\rm T}^{\gamma \gamma} &  {\rm T}^{e \gamma} & 
         {\rm T}^{p \gamma} &  \cdots \\
{\rm T}^{\gamma e} &  {\rm T}^{ee} &  {\rm T}^{pe} & \cdots \\
{\rm T}^{\gamma p} &  {\rm T}^{ep} &  {\rm T}^{pp} & \cdots \\
\vdots  & \vdots  & \vdots & \ddots \end{array} \right]
. 
\end{equation}

The transfer matrices depend on particle yields, $\rm
Y_{ij}^{\alpha\beta}$, which are defined as the probability of
producing a particle of type $\beta$ in the energy bin $j$ when a
primary particle of type $\alpha$ with energy in bin $i$
undergoes an interaction.  To calculate $\rm
Y_{ij}^{\alpha\beta}$ a Monte Carlo simulation can be used (see
Protheroe and Johnson (1995)\markcite{ProtheroeJohnson95} for
details).

%%%%%%%%%%%%%%%%%%%%%%%%%%%%%%%%%%%%%%%%%%%%%%%%%%%%%%%%%%%%%%%%%%%%%%%%%
\subsection{Matrix Doubling}
%%%%%%%%%%%%%%%%%%%%%%%%%%%%%%%%%%%%%%%%%%%%%%%%%%%%%%%%%%%%%%%%%%%%%%%%%
From Fig.~\ref{fig:e3kmfp}
we see that the smallest effective interaction length
is that for synchrotron losses by electrons at high energies.
We require $\delta x$ be much smaller than this distance which is
of the order of parsecs for the highest magnetic field considered.
To follow the cascade for a distance corresponding to a redshift
of $z \sim 9$, and to complete the calculation of the cascade 
using repeated application of the transfer matrices 
would require $ \sim  10^{12}$ steps.
This is clearly impractical, and one must use the more sophisticated approach
described below.

The matrix method and matrix doubling technique have been used
for many years in radiative transfer problems (van de Hulst 1963,
Hovenier 1971)\markcite{Hul63,Hov71}.  The method described by
Protheroe and Stanev (1993)\markcite{Pro93} is summarized below.
Once the transfer matrices have been calculated for a distance
$\delta x$, the transfer matrix for a distance $2\delta x$ is
simply given by applying the transfer matrices twice, i.e.
\begin{equation}
[{\rm T}(2 \delta x)] = [{\rm T}(\delta x)]^2.
\end{equation}
In practice, it is necessary to use high-precision during computation
(e.g. double-precision in FORTRAN), and to ensure
that energy conservation is preserved after each doubling.
The new matrices may then be used to
calculate the transfer matrices for distance $4\delta x$, $8\delta x$, 
and so on.
A distance $2^n\delta x$ only requires the application of this
`matrix doubling' $n$ times. 
The spectrum of electrons and photons after a large
distance $\Delta x$ is then given 
by
\begin{equation}
[{\rm F}(x +\Delta x)] = [{\rm T}(\Delta x)][{\rm F}(x)]
\end{equation}
where $[{\rm F}(x)]$ represents the input spectra, and $\Delta x=2^n\delta x$.
In this way, cascades over long distances can be modelled quickly and 
efficiently.

%%%%%%%%%%%%%%%%%%%%%%%%%%%%%%%%%%%%%%%%%%%%%%%%%%%%%%%%%%%%%%%%%%%%%%%%%%%%%%%

\section{The Origin of Cosmic Rays between 100 TeV  and 300 EeV}
%%%%%%%%%%%%%%%%%%%%%%%%%%%%%%%%%%%%%%%%%%%%%%%%%%%%%%%%%%%%%%%%%%%%%%%%%%%%%%%

The highest energy cosmic rays show no major differences in their
air shower characteristics to cosmic rays at lower energies.  One
would therefore expect the highest energy cosmic rays to be
protons particularly, if as is most likely, they are extragalactic
in origin.  However, it is possible that they are not single
nucleons.  Obvious candidates are heavier nuclei (e.g. Fe),
$\gamma$-rays and neutrinos.  In general it is even more
difficult to propagate nuclei than protons, because of the
additional photonuclear disintegration which occurs (Tkaczyk et
al. 1975, Puget et al. 1976, Karakula and Tkaczyk 1993, Elbert and
Sommers 1995, Anchordoqui et al.\ 1997, Stecker and Salamon  
1999)\markcite{Tka75,Pug76,KarakulaTkaczyk93,Elb95,Anchordoqui97,SteckerSalamon99}.
The possibility that the 300 EeV event is a $\gamma$-ray has been
discussed recently (Halzen et al. 1995)\markcite{Hal94} and,
although not completely ruled out, the air shower development
profile seems inconsistent with a $\gamma$-ray primary.  Weakly
interacting particles such as neutrinos will have no difficulty
in propagating over extragalactic distances, of course.  This
possibility has been considered, and generally discounted (Halzen
et al. 1995, Elbert and Sommers 1995)\markcite{Hal94,Elb95},
mainly because of the relative unlikelihood of a neutrino
interacting in the atmosphere, and the necessarily great increase
in the luminosity required of cosmic sources.  Magnetic monopoles
accelerated by magnetic fields in our Galaxy have also been
suggested (Kephart and Weiler 1996)\markcite{Kep95} and can not
be ruled out as the highest energy events until the expected air
shower development of a monopole-induced shower is worked out.

The subject of possible acceleration sites of cosmic rays at
these energies has been reviewed by Hillas (1984) and Axford
(1994)\markcite{Hil84,Axf94}, and one of the very few plausible
acceleration sites may be associated with the radio lobes of
powerful radio galaxies, either in the hot spots (Rachen and
Biermann 1993)\markcite{RachenBiermann93} or possibly the cocoon
or jet (Norman et al. 1995)\markcite{Nor95}.  One-shot processes
such as magnetic reconnection (e.g. in jets or accretion disks)
comprise another possible class of sources (Haswell et al. 1992,
Sorrell 1987)\markcite{Has92,Sor87}.

Acceleration at the termination shock of the galactic wind from
our Galaxy has also been suggested by Jokipii and Morfill
(1985)\markcite{Jok85}, but due to the lack of any statistically
significant anisotropy associated with the Galaxy it is unlikely to
be the explanation.  However, a recent re-evaluation of the
world data set of cosmic rays has shown that there is a
correlation of the arrival directions of cosmic rays above 40 EeV
with the supergalactic plane (Stanev et
al. 1995)\markcite{Sta95}, lending support to an extragalactic
origin above this energy, and in particular to models where
``local'' sources ($<100$ Mpc) would appear to cluster near the
supergalactic plane.  Such a correlation would also be consistent
with a Gamma Ray Burst (GRB) origin as several GRB have now been
identified with galaxies.

\subsection{Active Galactic Nuclei}

\begin{figure}[htb]
\plotone{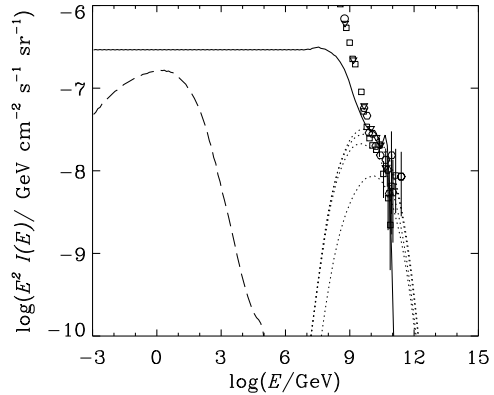}
\caption{Cosmic ray proton intensity multiplied by $E^{2}$ in the
model of Rachen and Biermann for $H_0=75$ km s$^{-1}$ Mpc$^{-1}$
with proton injection up to $3 \times 10^{11}$ GeV (solid line).
Also shown are intensities of neutrinos (dotted lines, $\nu_\mu ,
\bar{\nu}_\mu, \nu_e, \bar{\nu}_e$ from top to bottom), and
photons (long dashed lines).  Data are from Stanev
(1992)\protect\markcite{Stanev}; large crosses at EeV energies
are an estimate of the proton contribution to the total intensity
based on Fly's Eye observations.  (From Protheroe and Johnson
1995\protect \markcite{ProtheroeJohnson95}).
\label{fig:rachen_model}}
\end{figure}

Rachen and Biermann (1993)\markcite{RachenBiermann93} have
demonstrated that cosmic ray acceleration in Fanaroff-Riley Class
II radio galaxies can fit the observed spectral shape and the
normalization at 10 -- 100 EeV to within a factor of less than
10.  The predicted spectrum below this energy also fits the
proton spectrum inferred from Fly's Eye data (Rachen et
al. 1993)\markcite{Rac93a}.  Protheroe and Johnson
(1995)\markcite{ProtheroeJohnson95} have repeated Rachen and
Biermann's calculation to calculate the flux of diffuse neutrinos
and $\gamma$-rays which would accompany the UHE cosmic rays, and
their result is shown in Fig.~\ref{fig:rachen_model}.  The flux
of extremely high energy neutrinos may give important clues to
the origin of the UHE cosmic rays.  (For a review of high energy
neutrino astrophysics see Protheroe 1998\markcite{Protheroe98})

Very recently, Farrar and Biermann
(1998)\markcite{FarrarBiermann98} have found a remarkable
correlation between the arrival directions of the five highest
energy cosmic rays having well measured arrival directions and
radio-loud flat spectrum radio quasars with redshifts ranging
from 0.3 to 2.2.  The probability of obtaining the observed
correlation by chance is estimated to be 0.5\%.  Although the
statistical significance is not overwhelming, if further evidence
is provided for this correlation the consequences would be far
reaching.  The distances to these AGN are far in excess of the
energy-loss distance for pion photoproduction by protons.
Furthermore, given the existence of intergalactic magnetic
fields, any charged particle would be significantly deflected and
there should be no arrival direction correlation with objects at
such distances.  Hence, the particles responsible must be stable,
neutral, and have a very low cross section for interaction with
radiation.  Of currently known particles, only the neutrino fits
this description, however supersymmetric particles are another
possibility.  The difficulty of having high energy neutrinos
producing the highest energy cosmic rays directly is circumvented
if the neutrinos interact well before reaching Earth and produce
a particle or particles which will produce a normal looking air
shower.  Farrar and Biermann (1998) note that, as suggested by
Weiler (1998)\markcite{Weiler98}, this may occur due to
interactions with the 1.9~K cosmic background neutrinos.  Very
recent calculations by Yoshida et al
(1998)\markcite{YoshidaSiglLee98} have shown that as a result of
such interactions, and subsequent cascading, a flux of ultra high
energy $\gamma$-rays would result.  The clustering of massive relic
light neutrinos in hot dark matter galactic halos would give an
even denser nearby target for ultra high energy neutrinos as
suggested by Fargion et al. (1997)\markcite{FargionMeleSalis97}.

\subsection{Gamma ray bursts}

Gamma ray bursts (GRB) provide an alternative scenario for
producing the highest energy cosmic rays (Waxman 1995, Vietri
1995) \markcite{Waxman95,Vietri95}, and this possibility has
received renewed attention following the discovery that they are
cosmological.  GRB are observed to have non-thermal spectra with
photon energies extending to MeV energies, and in some GRB to
much higher energies (GRB~940217 was observed by EGRET up to 18
GeV, Hurley et al. 1994\markcite{Hurley94}).  Recent
identification of GRB with galaxies at large redshifts (e.g.
GRB~971214 at $z=3.42$, Kulkarni et al 1998
\markcite{Kulkarni98}) show that the energy output in
$\gamma$-rays alone from these objects can be as high as $3
\times 10^{53}$~erg if the emission is isotropic, making these
the most energetic events in the Universe.  GRB~980425 has been
tentatively identified by Galama et al (1998) \markcite{Galama98}
with an unusual supernova in ESO~184-G82 at a redshift of
$z=0.0085$ implying an energy output of $10^{52}$~erg.  These
high energy outputs, combined with the short duration and rapid
variability on time-scales of milliseconds, require highly
relativistic motion to allow the MeV photons to escape without
severe photon-photon pair production losses.  The energy sources
of GRB may be neutron star mergers with neutron stars or with
black holes, collapsars associated with supernova explosions of
very massive stars, hyper-accreting black holes, hypernovae,
etc. (see Popham et al 1998, Iwamoto et al 1998
\markcite{Popham98,Iwamoto98} for references to these models).

In the~relativistic~fireball model of GRB (Meszaros and Rees 1994
\markcite{MesRees}) a relativistic fireball sweeps up mass and
magnetic field, and electrons are energized by shock acceleration
and produce the MeV $\gamma$-rays by synchrotron radiation.
Protons will also be accelerated, and may interact with the MeV
$\gamma$-rays producing neutrinos via pion photoproduction and
subsequent decay at energies above $\sim 10^{14}$~eV (Waxman and
Bahcall 1997, Rachen and Meszaros 1998).
\markcite{WaxmanBahcall97,RachenMeszaros98}  Acceleration of protons
may also take place to energies above $10^{19}$~eV, producing a
burst of neutrinos at these energies by the same process (Vietri
1998)\markcite{Vietri98}.  These energetic protons may escape from
the host galaxy to become the highest energy cosmic rays (Waxman
1995, Vietri 1995)\markcite{Waxman95,Vietri95}.  Additional
neutrinos due to interactions of the highest energy cosmic rays
with the CMBR will be produced as discussed in the previous
section.

\subsection{Topological defects}

Finally, I discuss perhaps the most uncertain of the components
of the diffuse high energy neutrino background, that due to
topological defects (TD).  In a series of papers (Hill 1983,
Aharonian et al. 1992, Bhattacharjee et al. 1992, Gill and Kibble 1994)
\markcite{Hill83,AharonianBhatSchramm92,BhatHillSchramm92,GillKibble94},
TD have been suggested as an alternative explanation of the
highest energy cosmic rays.  In this scenario, the observed
cosmic rays are a result of top-down cascading, from somewhat
below (or even much below, depending on theory) the GUT scale
energy of $\sim 10^{16}$ GeV (Amaldi 1991)\markcite{Amaldi91},
down to $10^{11}$ GeV and lower energies.  These models put out
much of the energy in a very flat spectrum of neutrinos, photons
and electrons extending up to the mass of the ``X--particles''
emitted.

Protheroe and Stanev (1996)\markcite{ProtheroeStanev96} argued
that these models appear to be ruled out by the GeV $\gamma$-ray
intensity produced in cascades initiated by X-particle decay for
GUT scale X-particle masses.  The $\gamma$-rays result primarily
from synchrotron radiation of cascade electrons in the
extragalactic magnetic field and were found to just exceed the
observed diffuse $\gamma$-ray background for a magnetic field of
$10^{-9}$~G and X-particle mass of $1.3 \times 10^{14}$~GeV if
the observable particle intensity was normalized to the UHE
cosmic ray data.  Thus , for such magnetic fields and higher
X-particle masses (e.g. GUT scale), TD cannot explain the highest
energy cosmic rays.  Indeed there is evidence to suggest that
magnetic fields between galaxies in clusters could be as high as
$10^{-6}$~G (Kronberg 1994)\markcite{Kronberg94}.  However, for
lower magnetic fields and/or lower X-particle masses the TD
models might explain the highest energy cosmic rays without
exceeding the GeV $\gamma$-ray limit.  For example, Sigl et
al.~(1997)\markcite{SiglLeeSchrammCoppi96} show that a TD origin
is not ruled out if the extragalactic field is as low as
$10^{-12}$~G, and Birkel \&
Sarkar~(1998)\markcite{BirkelSarkar98} adopt an X-particle mass
of $10^{12}$~GeV.

I emphasize that the TD model predictions are {\em not} absolute
predictions, but the intensity of $\gamma$-rays and nucleons in
the resulting cascade is normalized in some way to the highest
energy cosmic ray data.  It is my opinion that GUT scale TD
models are neither necessary nor able to explain the highest
energy cosmic rays without violating the GeV $\gamma$-ray flux
observed.  The predicted neutrino intensities are therefore {\em
extremely} uncertain.  Nevertheless, it is important to search
for such emission because, if it is found, it would overturn our
current thinking on the origin of the highest energy cosmic rays
and, perhaps more importantly, our understanding of the Universe
itself.

%%%%%%%%%%%%%%%%%%%%%%%%%%%%%%%%%%%%%%%%%%%%%%%%%%%%%%%%%%%%%%%%%%%%%%%

\section{Observability of Ultra High Energy Gamma Rays}

In at least one of the origin models discussed above, a
significant fraction of the ``observable particles'' at 300 EeV
are $\gamma$-rays, and so it is appropriate to consider the
detectability of such energetic photons.  Above 100 EeV the
interaction properties of $\gamma$-rays in the terrestrial
environment are very uncertain.  Two effects may play a
significant role: interaction with the geomagnetic field, and the
Landau-Pomeranchuk-Migdal (LPM) effect (Landau and Pomeranchuk 1953,
Migdal 1956)\markcite{LPM1,LPM2}.

Energetic $\gamma$-rays entering the atmosphere will be subject
to the LPM effect (the suppression of electromagnetic
cross-sections at high energies) which becomes very important at
ultra-high energies.  The radiation length changes as $(E/E_{\rm
LPM})^{1/2}$, where $E_{\rm LPM} = 6.15 \times 10^4 \ell_{\rm
cm}$ GeV, and $\ell_{\rm cm}$ is the standard Bethe-Heitler
radiation length in cm (Stanev et
al. 1982)\markcite{StanevLPM82}.  Protheroe and Stanev
(1996)\markcite{ProtheroeStanev96} found that average shower
maximum will be reached below sea level for energies $5 \times
10^{11}$ GeV, $8 \times 10^{11}$ GeV, and $1.3 \times 10^{12}$
GeV for $\gamma$--rays entering the atmosphere at $\cos{\theta}$
= 1, 0.75, and 0.5 respectively.  Such showers would be very
difficult to reconstruct by experiments such as Fly's Eye and at
best would be assigned a lower energy.

Before entering the Earth's atmosphere $\gamma$-rays and electrons
are likely to interact on the geomagnetic field (see Erber
(1968)\markcite{Erber68} for a review of the theoretical and
experimental understanding of the interactions).  In such a case
the $\gamma$-rays propagating perpendicular to the geomagnetic field
lines would cascade in the geomagnetic field, i.e. pair
production followed by synchrotron radiation.  The cascade
process would degrade the $\gamma$-ray energies to some extent
(depending on pitch angle), and the atmospheric cascade would
then be generated by a bunch of $\gamma$-rays of lower energy.
Aharonian et al. (1991)\markcite{AharonianKanevskySahakian91}
have considered this possibility and conclude that this bunch
would appear as one air shower, having the energy of the initial
gamma-ray outside the geomagnetic field, being made up of the
superposition of many air showers of lower energy where the LPM
effect is negligible.  If this is the case, then $\gamma$-rays above
300 EeV would be observable by Fly's Eye, etc.  There is however,
some uncertainty as to whether pair production will take place in
the geomagnetic field.  This depends on whether the geomagnetic
field spatial dimension is larger than the formation length of
the electron pair, i.e. the length required to achieve a
separation between the two electrons that is greater than the
classical radius of the electron (see also Stanev and Vankov
1996)\markcite{StanevVankov96}.

%%%%%%%%%%%%%%%%%%%%%%%%%%%%%%%%%%%%%%%%%%%%%%%%%%%%%%%%%%%%%%%%%%%%%%%%%%%%%

\section{Affect of large-scale magnetic structure}

At low energies, propagation of cosmic rays through the Galaxy
smears out their arrival directions, giving rise to an almost
isotropic distribution.  However, simulations of cosmic rays
through a model of the galactic magnetic field including a
turbulent component shows that for energies greater than $\sim
$Z~EeV significant anisotropies may be expected from sources in
our Galaxy (see, e.g., Lee and Clay 1995\markcite{LeeClay95}).
Clearly, at much higher energies the magnetic field of our galaxy
plays an insignificant role in smearing directions of
extragalactic cosmic ray nuclei.  Ultra-high energy cosmic rays
can be subject, however, to significant deflection and energy
dependent time delays in large scale extragalactic or halo
magnetic fields.  Lemoine et al. (1997)\markcite{Lemoine97} have
performed Monte Carlo simulations of propagation which show how the
duration of cosmic ray emission in distant sources has a dramatic
effect on the observed spectrum.  For example, in one model, a
bursting source at 30~Mpc with an $E^{-2}$ spectrum gave a spectrum
very strongly peaked at 110 EeV.

Ryu et al. (1998)\markcite{RyuKangBiermann98} consider the
possibility that the cosmic magnetic field, instead of being
uniformly distributed, is strongly correlated with the large
scale structure of the Universe.  They derive an upper limit to
the magnetic field in filaments and sheets of $1 \mu G$ which is
$\sim10^3$ times higher than the previously quoted values.
Clearly, if such cosmic structures have high magnetic fields this
will have important consequences for cosmic ray acceleration and
propagation.  For example, Tanco (1998b)\markcite{Tanco98b} has
considered propagation through well ordered compressed magnetic
fields ($\sim 0.1 \mu$G) inside cosmological walls and found the
cosmic ray flux leaving a wall to be highly anisotropic (by 2 or
3 orders of magnitude), being greatest near the central
perpendicular of the wall.  Thus, we might expect an anisotropy
in the direction perpendicular to the nearest cosmological wall
(Centaurus).  He also found that arrival directions of CR are
smeared, and correlations between $\gamma$-rays and CR in bursting
events are lost.

Sigl, Lemoine and Biermann (1998)\markcite{SiglLemoineBiermann98}
have considered propagation in the local supercluster with
$B_{\rm rms} \sim 0.1 \mu$G, with a maximum eddy length of 10~Mpc
and a distance to the nearest source of 10~Mpc.  For a soft
injection spectrum, $E^{-2.4}$, and with both the source and the
observer in the local supercluster (sheet with thickness 10~Mpc)
they found excellent agreement with the spectrum observed above
10 EeV.  Below 100 EeV cosmic rays diffuse, with arrival
directions at Earth being spread out over $\sim 80^\circ$ while
above 200 EeV they are just deflected but still have a
significant spread in arrival directions of $\sim 10^\circ$
explaining the lack of correlation of the highest energy cosmic
ray with any known nearby source.  Tanco
(1998a)\markcite{Tanco98a} noted that the local distribution of
luminous matter is far from uniform, and using a redshift
distribution based on the CfA redshift catalog they found
agreement with the observed super-GZK spectrum.  A similar
conclusion was reached by Blasi and Olinto
(1998)\markcite{BlasiOlinto98}.

%%%%%%%%%%%%%%%%%%%%%%%%%%%%%%%%%%%%%%%%%%%%%%%%%%%%%%%%%%%%%%%%%%%%%%%%%%%%%

\section{Conclusion}

While the origin of the highest energy cosmic rays remains
uncertain, there appears to be no necessity to invoke exotic
models.  Shock acceleration, which is believed to be responsible
for the cosmic rays up to at least 100 GeV, is a well-understood
mechanism and there is evidence of shock acceleration taking
place in radio galaxies (Biermann and Strittmatter
1987)\markcite{BS87}, and the conditions there are favourable for
acceleration to at least 300 EeV.  Such a model, in which cosmic
rays are accelerated in Fanaroff-Riley Class II radio galaxies,
can readily account for the flat component of cosmic rays which
dominates the spectrum above $\sim 10$ EeV (Rachen et
al. 1993)\markcite{Rac93a}.  Indeed, one of the brightest FR II
galaxies, 3C 134, is a candidate source for the 300 EeV Fly's Eye
event (Biermann, 1997)\markcite{Biermann97}.

Whatever the source of the highest energy cosmic rays, because of
their interactions with the radiation and magnetic fields in the
Universe, the cosmic rays reaching Earth will have spectra,
composition and arrival directions affected by propagation.  In
particular the composition of the arriving particles may differ
drasticly from that on acceleration.  For example, the
accompanying fluxes of neutrinos produced in the sources and by
cascade processes during propagation to Earth may dominate the
spectrum at highest energies.  The astrophysics ingredients to
the propagation (e.g., magnetic field intensities and
configurations) are still somewhat uncertain, and we need better
statistics on the arrival directions, energy spectra and
composition, as well as the intensity of the diffuse backgrounds
of very high energy neutrinos and $\gamma$-rays (partly produced in
cascades initiated by cosmic ray interactions).  

The Auger Project (Auger Collaboration 1995), an international
collaboration to build two UHE cosmic ray detectors, one in the
United States and one in Argentina, each having a collecting area
of about 3000 km$^2$ \markcite{AUGER}, and will measure the
energy spectrum and anisotropy of the highest energy cosmic rays
with the required precision, and will also be sensitive to
extremely high energy neutrinos (Capelle et
al. 1998)\markcite{Capelle98}.  The next generation neutrino
telescopes, such as the planned extension of
AMANDA\markcite{Halzen98}, ICECUBE (Shi et al
1998)\markcite{ICECUBE}, and ANTARES (Blanc et
al. 1997)\markcite{ANTARES}, may have effective areas of
0.1~km$^3$, or larger, and be sufficiently sensitive to detect
bursts of neutrinos from extragalactic objects and to map out the
spectrum of the diffuse high energy neutrino background.

Once we have better better statistics of particle fluxes we will
be better placed to understand the origin of the highest energy
particles occurring in nature.  Furthermore, once the
astrophysics is worked out, cosmic rays will be the tool for
exploring particle physics well above terrestrial accelerator
energies.

\section*{Acknowledgments}
This chapter is based in part on a lecture given at Erice
(Protheroe 1996\markcite{Protheroe96}).  I thank Qinghuan Luo,
Anita M\"{u}cke and Peter Biermann for reading the manuscript.
My research is supported by a grant from the Australian Research
Council.

\end{document}